# Analysis of Phishing Attacks and Countermeasures

Biju Issac, Raymond Chiong and Seibu Mary Jacob
Information Security Research Lab, Swinburne University of Technology, Kuching, Malaysia
{bissac, rchiong, sjacob}@swinburne.edu.my

**Abstract**
One of the biggest problems with the Internet technology is the unwanted spam emails. The well-disguised phishing email comes in as part of the *spam and makes its entry into one's inbox quite* frequently nowadays. While phishing is normally considered a consumer issue, the fraudulent tactics the phishers use are now intimidating the corporate sector as well. In this paper, we analyze the various aspects of phishing attacks and draw on some possible defenses as countermeasures. We initially address the different forms of phishing attacks in theory, and then look at some examples of attacks in practice, along with their common defenses. We also highlight some recent statistical data on phishing scam to project the seriousness of the problem. Finally, some specific phishing countermeasures at both the user level and the organization level are listed, and a multi-layered anti-phishing proposal is presented to round up our studies.

## 1. Introduction

As technology advances, the Internet along with email has become an integral part of one's life. Unfortunately, the flexibility provided by the advancement of technology has at the same time resulted in criminals following the trend. Many problems thus arise, and one of such is the identity theft. Recently, one form of identity theft crime that has become a lethal security threat is phishing, targeted primarily at the casual email users. Phishing is the act of sending forged emails and fake websites to users in an attempt to scam them into surrendering personal information that leads to identity theft [1]. Typical phishing email is sent to many potential victim's mailboxes, and usually comes with a clickable link. It is meant to lure the recipient to trust that the email received is from a trusted source so that the recipient will open and click on the provided website hyperlink which connects them to some fake websites, and eventually extracting personal information from them. The phishers (attackers) may use deceptive sender address, genuine-looking logo and fraudulent web links in such emails. In the battle against phishing, we would like to emphasize the fact that user education is important, as ignorant users can get themselves into troubles even with the best and most sophisticated defenses available.

The rest of this paper is organized as follows. Section 2 explains the different types of phishing attacks in theory. Following which, section 3 discusses some examples of phishing attacks as well as their corresponding defenses in practice. Section 4 shows some recent statistical details on the phishing scams. Section 5 explains on specific countermeasures and section 6 presents a multi-layered anti-phishing proposal. Finally, section 7 concludes this paper with summary and remarks.

## 2. Types of Phishing Attacks

Phishing attacks target mostly on confidential information such as user names, passwords, social security numbers, passport numbers, credit card numbers, bank account numbers, PIN numbers, birthdates, mother's maiden names, etc. Phishers can easily focus on the technology expertise and sit in the comfort of their homes or hack offices to get sensitive information at their fingertips. In this section, we would like to discuss several types of phishing attacks as listed below, which were discussed by Emigh [2]:

1. Phishing Attack by Fraud, where the user is fooled by fraudulent emails to disclose personal or confidential information.
2. Phishing Attack by Infectious software, where the attacker succeeds in running dangerous software on user's computer.
3. Phishing Attack by DNS spoofing, where the attacker compromises the domain lookup process so that the user's click would lead him or her to a fake website.
4. Phishing Attack by Inserting harmful content, where the attacker puts malicious content into a normal website.
5. Phishing Attack by MITM approach, where the attacker gets in between the user and the legitimate site and taps sensitive information.
6. Phishing Attack by Search Engine indexing, where the fake web pages with attractive offers created by the attacker gets indexed by a search engine, so that a user would stumble upon it.

In phishing attack by fraud, the attacker sends a fraud email demanding the user to take some action, normally by citing a problem with his bank account, advertising a new service roll-out, or offering fictitious invoice, etc. In all the above cases, the user is directed to a website where one's personal or sensitive information is being extracted. The attacker may use a link with a domain name that looks very similar to the original domain name. If responded positively, this can lead to malicious software being installed on the user's computer which leaves an open back door for future attacks. In phishing attack by infectious software, the attacker takes advantage of security vulnerabilities with the computer or the operating system. Often, it happens by luring the user to open an email attachment with promise of pornographic images or other interesting baits. Some freely downloadable software also contains infectious programs that get installed along with the original software. Keyloggers are stealth programs that can be installed into a web browser and/or work as device driver that captures the data that is keyed in by the user and sent to a remote server setup by the attacker. Session hijacking can also happen through a malicious browser component that was installed by the attacker. When the user logs in to do a transaction, the infectious software hijacks that session and does malicious activity once the user credentials are proved to be right with the transacting website. Web Trojans that pop up to collect user credentials and channel them back to the attacker are also prevalent nowadays. In phishing attacks by DNS spoofing, the DNS lookup process is compromised either on the local computer or the DNS server. Hosts file in a local computer is looked into first before querying a DNS server to find the IP address to domain mapping when a link is clicked or when a domain address is entered in the browser. If this file is compromised and false mapping is entered in hosts file through malicious software, the user can ignorantly go to the attacker's website and give personal information. A Crowt.D worm attack in year 2005 was doing this. System configuration altering attacks can be done to compromise the DNS server, so that the mapping is poisoned. In phishing attacks by inserting harmful content, the attacker can compromise a server's security vulnerability and put malicious or harmful content instead of legitimate one, such as the cross-site scripting (XSS) vulnerability. Here content coming from external sources like chat message, search item or web email would be supplied to the visitor's web browser. SQL injection vulnerability can also be used to perform malicious actions.

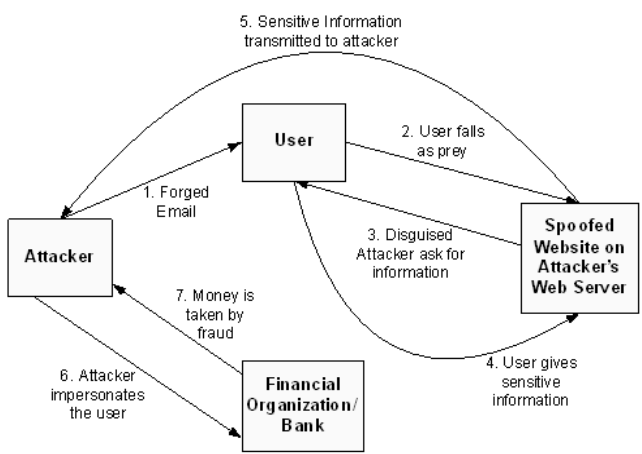

Fig 1. Stages in a Phishing attack

In phishing attacks by MITM approach, the attacker intercepts user traffic by standing between him and the website. He uses proper response forwarding mechanism as the user communicates with the intended site and helps communication back to the user from the website – all through his computer. The user thus can have no suspicion on traffic tapping. In phishing attacks by search engine indexing, the attacker creates genuine-looking website for fake products where he has options to perform financial transactions and attract users because of better offers than others. This website would then be submitted for indexing by search engines, so that any user can get a hit on the attacker's webpage. Fraudulent banks with higher interest rates can attract customers in such a scenario and make the users to perform some cash transfer to the newly created account in the attacker's web trap [2]. The stages of a general phishing attack are summarized in figure 1.

### 3. Examples of Phishing Attacks and Defenses

In this section, we describe the technical aspects of typical phishing attacks based on Beardsley's paper [1]. Four common techniques are discussed, namely email field manipulation, email with image-only content, misdirection and redirection, and pop-up window attack. The first two techniques, email field manipulation and email with image-only content, can be categorized under type 1 to 3 as discussed in the previous section. Meanwhile, misdirection and redirection as well as pop-up window attack can be categorized under type 1 to 6. We have also briefly outlined the corresponding defenses that could prevent the phishing expedition and protect consumers from the online fraud

Attack 1: Email Field Manipulation attack
One of the most common phishing techniques comes in forging the email headers. As most of the email client software depends heavily on the message's From field to determine a sender, many phishing emails simply forge the From message's header. It is quick and literally requires no tough effort to forge the From field, and the technique of doing so is widely known since the 1980s. A forged From field can easily be configured on an SMTP server that allows the use of it without authentication. To make the matter worse, some anti-spam client applications even allow the use of emails that merely match the From patterns, thus bypassing the last line of anti-spam defense. Recently, more sophisticated phishing emails alter not just the From field but the Received path headers too [1]. In figure 2, the E-mail header of a phishing email from phisher@hotmail.com (Phisher) to victim_user@target.com (Victim User) disguised as clean_user@yahoo.com (Clean User) is shown.

----------------------------------------------------------------
Return-Path: <phisher@hotmail.com>
Received: from hotmail.com (bay20-dav5.bay20. hotmail.com [64.4.54.185]) by target.com (8.12.11.20060308/8.12.11) with SMTP id k4L8ighF012529 for <victim_user@target.com>; Sun, 21 May 2006 16:44:43 +0800
Received: from mail pickup service by hotmail.com with Microsoft SMTPSVC; Sun, 21 May 2006

```
Message-ID: <BAY20-DAV5625D82C250389FF8
32A8B0A50@phx.gbl>
Received:  from  60.1.120.150  by  BAY20-
DAV5.phx.gbl with DAV; Sun, 21 May 2006
X-Originating-IP: [60.1.120.150]
X-Originating-Email: [phisher@hotmail.com]
X-Sender: phisher@hotmail.com
Reply-To: <clean_user@yahoo.com>
From: "Clean User" <clean_user@yahoo.com>
To: "Victim User" <victim_user@target.com>
Subject: Great news ... Click the link
Date: Sun, 21 May 2006 17:02:38 +0800
X-Mailer: Microsoft Office Outlook 11
------------------------------------------------------------------
```

Fig 2. The E-mail header of a phishing email from Hotmail Server

We have used the hotmail SMTP server to perform the attack above. It was easily configured on MS Outlook. As hotmail's SMTP server does not use any kind of authentication, the phisher can disguise himself as virtually "anyone" to send phishing email to the victim user. If he disguises himself as the CEO or Manager of an organization and sends emails requesting for some personal information from the employees of the organization (with or without a forged web link), it would become an ultimate attack where every sincere employee would give out information to their boss! It is true that the header of such emails can reveal information that is suspicious as shown in figure 2, but the problem is how many people will actually check the header of an email? A more effective disguised attack can be done using free Mass eMailer ver.2.2 software or other similar ones, as shown in figure 3.

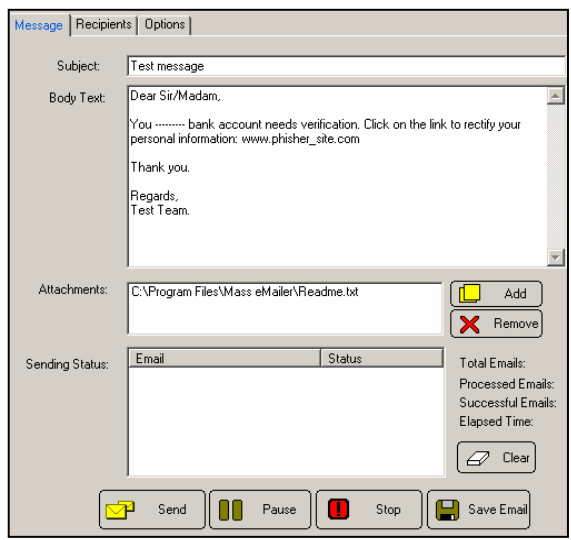

Fig 3. The Mass eMailer 2.2 software that can be used for sending mass mails with forged identity

Common Defense: Before we mention about the defense on forging fields in email attack, it is necessary to note that a phishing email received in general goes through a path that differs from a regular email slightly. It takes a detour from a trusted financial institution to some home broadband machine, and then finally arrives at the user's mail server. Hence, the user will see only the trustworthy Received header as it is the last hop before a trusted mail server. Although there is no sound authentication to tackle the forging of From fields, still certain tracks can be followed to sniff a forgery. As phishers use some random portions of the name and subject of the email message, the presence of strange spacing and characters may flag a phishing message. The combination of hash-busting characters with particular brand strings gives a strong indication that the email carries malicious intent. By comparing the timestamps included in Received headers, the more sophisticated forging emails can also be detected. If the Received chain is crafted, the time zones are normally blank, or the timestamps not contiguous.

Attack 2: Email with Image-only Content attack
Another common phishing technique is called the image insertion attack in emails [1]. It is a situation where the phishing emails contain only binary images, such as the .GIF pictures, in an HTML-capable mail client. The purpose is to supply a text stream of random words and phrases to confuse spam filters, thus bypassing them. Common Defense: The most straight forward solution for the image insertion attack is obviously to allow only the plain-text emails and disable HTML-based messages. However, we reckon that graphical appearance is one of the significant features the Internet users are expecting. As such, some more creative prevention is necessary for commercial purposes. Instead of totally disallowing the HTML-based emails, an alternative can be provided for the email users to choose between plain-text and HTML. If a certain email looks suspicious, the email user can always opt for the plain-text version to avoid being snared into some fraudulent web links.

Attack 3: Misdirection and Redirection attack
The misdirection and redirection attack is considered by many as one of the most creative phishing techniques. This attack misdirects the users through emailed URLs to fraudulent sites [1], [3]. One of the simplest ways to misdirect email users to unreliable sites is through the crafted links that can be generated automatically. Most of the mail clients available in the market nowadays instinctively convert plain-text http links into clickable URLs. The rendering link may appear to be a familiar-looking link such as https://logon.rhbbank.com.my/ with added security measure due to the apparent use of https, but in actual fact it is a mere illusion that misdirects the users to another site over http. For instance, <a href="http://192.9.200.1">https://logon.rhbbank.com.my/</a> will direct a user to an unknown site with IP address of http://192.9.200.1 instead of the trusted https site, even though the URL appears to the user is a familiar-looking web link like https://logon.rhbbank.com.my/. Phishers also use overlapping area map tags to intersect two links in same clickable area within the phishing email. While the hovering link appears to be normal, it actually takes the user to the hidden link. Apart from the misdirection tactics, recently many phishers have been making full use of the available technologies and services offered by various well-known vendors or sources to redirect email users to some nested links [1].

Among other redirection tactics, some phishing emails are using the address obfuscation service to provide a shorter version clickable link in the email based on purposefully generated long URL. With this, the phishers are hoping that the email users will try the shorter URL rather than the longer one, and in turn will lead the users to the phisher's sites. The address obfuscation service is ideal for the phishers to hide chained redirects. Two other services that have become the phishers' instruments to perform identity theft are the free DNS redirectors and the embedded html forms. In free DNS redirectors, the phishers have plenty of ground to resolve the IP addresses to the chosen DNS names transparent to the email users. This can easily obfuscate the users to click on the fraudulent links. Meanwhile, embedded html forms are normally embedded within a phishing email, together with a typical call-to-action that requires the user to verify his or her account activity. This substitutes the requirement for the phishers to use a live Web server to extract user's private information. Common Defense: Fencing off phishing emails consisting of encoded or redirection tricks can be very straight forward with several easy-to-implement anti-phishing rules such as no hex-encoded printable ASCII characters in domain names, no HTTP link containing 'http' more than once and no nested <A> and <AREA> links [1]. The redirection with address obfuscation service, however, is more difficult to combat due to the fact that it is not strongly associated with phishing activities. By performing a matching and blocking based on the From header of known financial institutions with their URLs may be a way to tackle it.

Attack 4: Pop-up window attack
Pop-up login screens are used extensively by some of the commercial banks to perform basic login and customer tracking functions. However, in order for the users to login to their personal accounts, they have to disable the pop-up blocker provided by the Internet browser. In such situation, the phisher's site can immediately spawn two pop-up windows once the login is connected, with one window showing the real site, and the other the phisher's own pop-up window that asks only for identity information. Common Defense: User education is the best approach here, where the users can be advised to be alert to unexpected pop-ups, which can 'look' slightly different from the normal.

## 4. Phishing Scam Statistics and Analysis

The summary of Phishing Activity Trends Report (March 2006) by Anti-Phishing Working group (APWG) as given in [4] is shown in figure 4 below. The monthly Phishing Activity Trends Report analyzes the phishing attacks reported to the APWG through the website at http://www.antiphishing.org or through user email submission to APWG's email at reportphishing@antiphishing.org. The APWG phishing attack repository is quite revealing in terms of email fraud and phishing activity. The summary of findings in figure 4 indicates that phishing is a serious problem that is growing vigorously. The total number of unique phishing reports submitted to APWG in March 2006 was 18,480 – the highest recorded figure (see figure 5). Most phishing attacks and new phishing sites also occurred in the same month. The total number of unique phishing websites recorded by APWG in March 2006 was 9666, which is a continual increase since January 2006.

----------------------------------------------------------------
Number of unique phishing reports received in March 2006: 18,480
Number of unique phishing sites received in March 2006: 9666
Number of brands hijacked by phishing campaigns in March 2006: 70
Number of brands comprising the top 80% of phishing campaigns in March 2006: 3
Country hosting the most phishing websites in March 2006: United States
Contain some form of target name in URL: 48.05 %
No hostname, just IP address: 32 %
Percentage of sites not using port 80: 3.9 %
Average time online for site: 5.0 days
Longest time online for site: 31 days
----------------------------------------------------------------

Fig 4. Summary of March 2006 Phishing Report Findings [4]

In March 2006, the Websense Security Labs reported that the United States was placed at the first spot in the list for hosting phishing sites, with a 35.13% share.

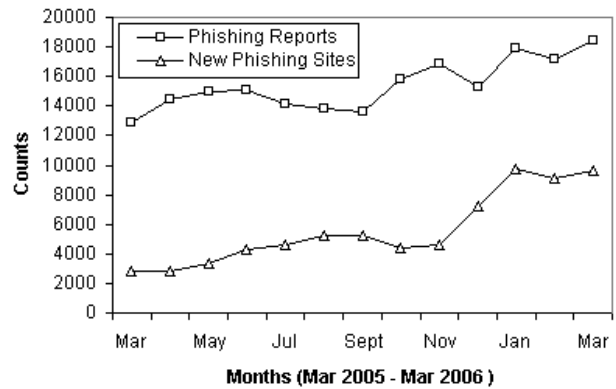

Fig 5. The line graphs that show the phishing report counts and new phishing sites counts [4].

The rest of the top 10 countries with their percentage contribution were as follows: China with 11.93%, Korea with 8.85%, Germany with 3.57%, Canada with 3.52%, Japan with 2.39%, Romania with 2.29%, Spain with 2.13%, Brazil with 1.97%, and Argentina with 1.92%. There was also a great increase in phishing based Trojans and keyloggers as shown in figure 6. The number of hijacked brand was also averaging around 91.23 from March 2005 to March 2006, as depicted in figure 7 [4].

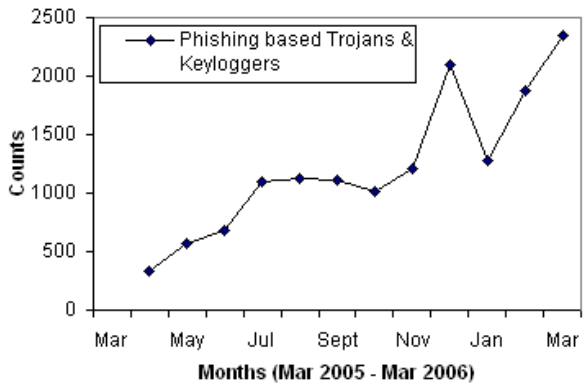

Fig 6. Line graph that show the phishing based Trojans and keyloggers counts [4]

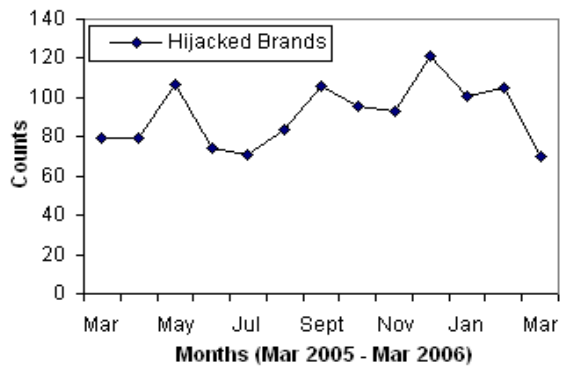

Fig 7. Line graph that shows the hijacked brand counts [4]

To have a rough idea on identity theft in general, we took a closer look at the Consumer Fraud and Identity Theft Complaint Data (January-December 2005). The report is registered by Federal Trade Commission (in the United States) and taken from the Consumer Sentinel and the Identity Theft Data Clearinghouse [5]. It gives details of identity thefts over the years 2003-2005. Checking against the reported amount paid by the consumers, the percentage of complaints reporting above $1000 payments topped at 23%, close to a 10% increase against the complaints in 2003 and 2004. As for the methods of payments reported by the consumers, there has been a sound increase from 5% in 2003 to roughly 20% in 2005. The credit card method adopted cases top the list and keep a consistent 40% over the 3 years. Coming to the question of which age group seems most vulnerable to the above fraud cases, the age group between 20-50 years accounts for a total of 75% throughout the years, with each age group between 20-29 years, 30-39 years and 40-49 years equally distributed among that 75%. As for the identity theft complaints, the age group between 18-30 years is the most victimized among the lot [5].

**5. Countermeasures against Phishing Scam**
There are some steps that users or organizations can take when it comes to handling spam and phishing emails. Some general precautions that can be taken against phishing at a user level are as follows [6]:

1. Within a phishing email, never click on hyperlinks, as they may not take the user to the right place.
2. Use proper and updated security software – like antivirus software, anti-spam software and anti-spyware software. Update the signatures of these software programs on a regular basis. This can harden the user computer against attacks. Even though anti-spam software may generate false alarms, it will be good to flag the suspicious emails and store them at some location (locally) for a quick review, rather than deleting them straightaway. Keep the browser and the operating system software updated with all the recent and critical security updates. A security update is released when there are critical security flaws that an attacker can utilize.
3. Never download free software from unknown websites as some free software downloads come with hidden or embedded Trojans.
4. Use a strong firewall, which is in-built in many operating systems nowadays.
5. Look for "https" protocol in the address bar and the padlock symbol on the bottom status bar of the browser, before entering any personal information.
6. Educate oneself of the fraudulent activities on the Internet and keep an updated knowledge of such happenings.
7. Anti-Phising toolbars can be used along with the Internet browsers to identify phishing sites and warn the user as he or she browses, if the site is suspicious [7].

There are some more detailed anti-phishing steps that an organization can consider. Some detailed countermeasures against phishing attack at an organization level are stated as follows [2], [8]:

1. Email filtering can be a good option to tackle spam or phishing emails. As phishing emails are a subset of spam, good spam filters can help. Signature based anti-spam filters could prove to be useful too. To validate any anti-spam software installation, the following can be done. Arrangements need to be done to only tag the spam initially. This is the first stage of implementation. After successful tagging, the spam could be directed to some location (Junk Email Store) on E-mail server where it can be monitored on a weekly basis by the email administrator. The false positives, if any, could then be directed to the right person. This is the second stage of implementation. The flow chart of such an arrangement is shown in figure 8.

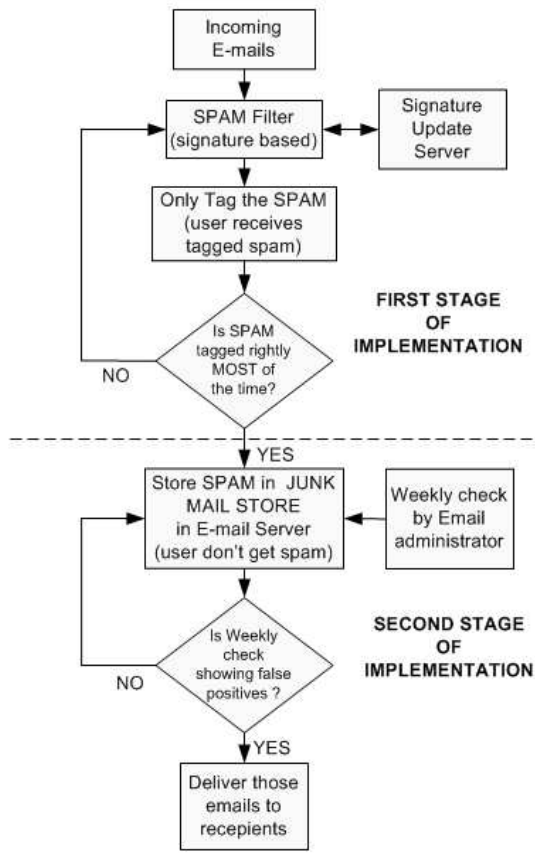

Fig 8. The flow chart showing Email filtering implementation and its validation

2. Email authentication can ensure that message is sent by the intended person who is the sender of the mail. The attacker normally forges the return address and would send email from a similar-looking domain to that of an original domain. There are different approaches proposed for email authentication, as of now. Return address forgery can be tackled by Sender-ID and SPF by checking DNS records to ensure whether the IP address of the sending MTA (Mail Transfer Agent) is an authorized sender. Domain level cryptographic signatures can also be used to provide authentication through Domain keys by cross-checking the DNS record. Cryptographically signed emails can be a good option especially if signing becomes a normal way of sending emails. Signing can be done either at the client machine or at the gateway.

3. Patching the operating system for security vulnerabilities can be a hardening measure to withstand malicious software's execution on user computer. Security updates when available can be downloaded to a local server and distributed through local area network.

4. User education is an important step in thwarting phishing scams. Users need to be instructed not to click on links in an email and to ensure that SSL is being used for secure transactions on web pages. They should be instructed to always use expected domain name for logging in. Also, they should never give in to a call-to-action in any email that warns of negative consequences if they fail to follow the link.

## 6. Multi-Layered Anti-Phishing Proposal

In this section we would like to propose a multi-layered approach to curb phishing emails. The approach is a combination of different technologies. The steps/stages can be combined in a suitable order into one anti-phishing solution which can be installed on the email server. A router based implementation of some modules is also a possibility.

Step 1: White List and Black List Approach: Implement a white-list, which is a list of "fully permitted" email addresses. For example, in a big organization where the email users are using a client interface, they can be allowed to submit the .csv format of their address book (without unwanted entries) through a simple upload to email server, which updates the central white-list for the organization. Also, once a spam escapes and comes into a user's mail box, through a one-button-press on email client, they should be able to report the mail as spam to their email server, which can improve the black-list entries. Their signatures can also be noted. When a reasonable number (say, n number) of people reports it as spam, the message can be marked as "spam" in users' mailboxes so that it is not delivered to them again [2]. Black-listed email addresses will also be ranked based on how many people reported it as spam or phishing address. Provisions can be made to block addresses based on domain names. Also, have all previous phishing signatures stored in a central online database. Details like email address, IP address, subject details, random lines from messages and keywords can be stored too. The server should contain all the spoofed company's original web pages that would be updated regularly. These web pages are used for layout and content comparison. All the entries in the database which are potential phishing sources are assigned a score, based on their frequency of being noted and caught. Their severity level would increase as the score increases. The setup is shown in figure 9.

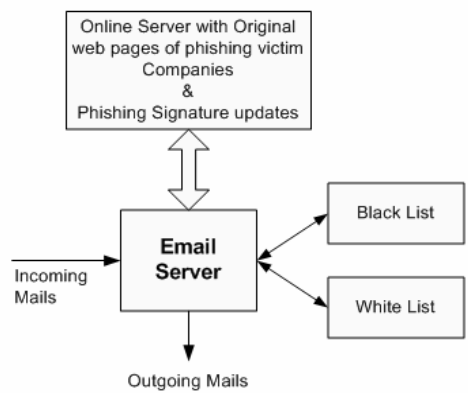

Fig 9. The White List/Black List with Online Server

Step 2: Securing of SMTP Server: SMTP servers from registered static IP address only should be allowed. It should support SMTP user authentication and be standardized to work only in this way. No SMTP relays should be allowed. SMTP servers should not be allowed to run from a dynamic IP address, as phishers could run their own SMTP servers from dial-up connections. Optionally, digital signatures can be gradually made mandatory in emailing systems so that sender identity cannot be forged. This will prevent further email messages with spoofed sender addresses as such emails would be rejected. Only a valid sender can now send emails.

Step 3: Implement Grey Listing: The Grey listing approach [9] proposed by Harris looks at three pieces of information that form a signature – the IP address of the host attempting the delivery, the envelope sender address and the envelope recipient address. If the receiving side has never seen this signature, the email would be rejected for the first time and it would become a bounced email. It would be allowed in only a second time (when the sender resends), after a delay of 25 minutes to 4 hours. Generally, this would stop spam and phishing emails to a great extent, since spammers and phishers may not resend (most of the time) their emails with the same signature. Some fine-tuning can be done to the above procedure by allowing a growing safe-list on the receiving email server that takes priority over grey listing. This can significantly reduce the bounced emails.

Step 4: Webpage Layout Comparison: When a new phishing email arrives which is sent through a non-spoofed email address, the web link in it would be compared with the web pages of known companies that are prone to spoofing (from past history), which are already stored in a central server database. The two web pages are checked for similarity on a 5-region wise comparison – south, north, west, east and center regions. Font attributes (like font type, style, size along with formatting attributes like bold, italics and underlining, etc.) and image related attributes (like image type, size, width, height, creation date, etc.) are checked in each region in both pages. If the comparison yields positive similarities result, the email would be accepted, otherwise the user is warned and the email's signature will be stored in the central online server. Scores can be assigned to comparison variables used and a decision can be taken based on the cumulative score crossing a threshold value. A very similar approach has already been proposed by Wenyin et. al. [10].

Step 5: Text Extraction from Image: When the email comes in with images as contents, the text in it would be extracted and used for lexical analysis using Bayesian filters (as in step 7), and the webpage that it is pointing to would be checked against a regular phishing email. Any discrepancy can thus be notified to the user and alerted to the central online server by storing its signature.

Step 6: Matching DNS names: The web links in phishing emails are also checked for veracity with the original organizations web domain, through a DNS query. If it is a concocted website link and a domain (that has nothing to do with advertised organization), the link can immediately be notified to the user and the central server database can be updated with the details. For example, consider a phishing email with Citibank details, asking the user to click a web link to update Citibank account details. The first 2 octets in IP address of Citibank in decimal dot notation is 192.193 and this can be checked with the forged domain's IP address.

Step 7: Implementing Aggregate Bayesian Filters: Bayesian filtering [11], [12] works on the principle that the probability of an event occurring in the future can be inferred from the previous occurrences of that event. For example, if the word "sexy" occurs in 1000 out of 5000 spam emails and 100 out of 500 legitimate emails, then the spam probability would be (1000/5000) / (1000/5000 + 100/500) = 0.5. Like spam filters, phishing emails can also be processed in the same way and be filtered through Bayesian filters using single keyword, 2-keyword or 3-keyword combinations. These can be three blocks with increasing priorities of scores and a weighted average score can be taken for labeling it as spam. This single keyword (say like – porn, horny), 2 keywords (like – porn site, horny wives) or 3 keywords (like – free porn site, sexy horny wives, etc.) self-learning filters can help to block suspicious spam emails. Keywords in subject line and content space can give different scores. Subject line keywords are crucial and will have high scores. Special characters (like $, -, *, digits 1-9, ', ", etc.) introduced by spammers to confuse spam filters can be extracted/removed or replaced (say, 0 with o) from keywords to improve filtering. If 2-keyword or 3-keyword combinations are not possible, only a single keyword will be considered and scored without weights. The score from 1-keyword block (i.e. score 1) has weight of X, the score from 2-keyword block (i.e. score 2) has weight of Y and the score from 3-keyword block (i.e. score 3) has weight of Z. We are proposing to take a weighted average ($W_{avg}$) which is given as follows: $W_{avg}$ = (X*score 1 + Y*score 2 + Z*score 3)/3, where X < Y and Y < Z. The value of X, Y and Z have to be fine-tuned and the best possible values should be chosen. If there can only be single and 2-keyword option, then different values of weights (X and Y, where X < Y) can be considered for their respective scores. A threshold value can be set and emails with weighted average values above the threshold will be tagged, quarantined or rejected. The same technology can be extended to phishing emails – with phishing related keywords (see figure 10 for details).

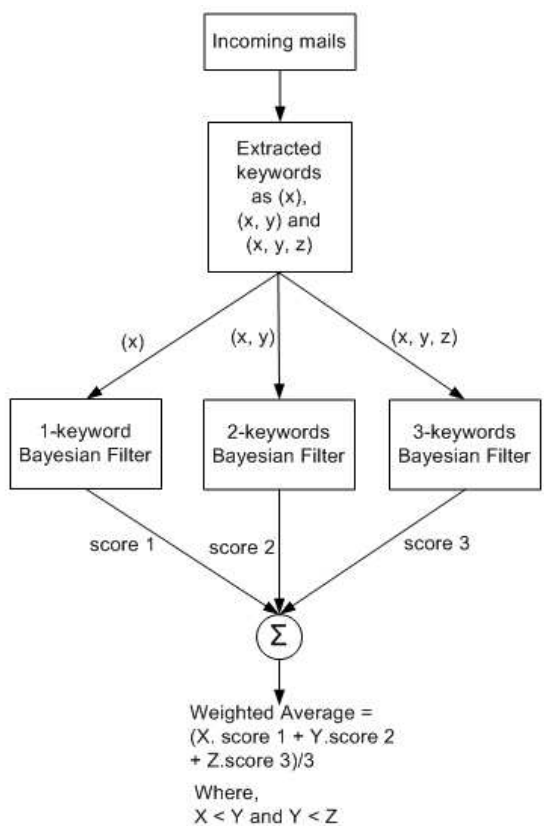

Fig 10. The aggregate Bayesian filtering approach

The above stages once combined, can be tested for accuracy of the phishing email filtering by initially tagging the phishing emails, as explained in figure 8. Threshold values can also be fine tuned in different stages.

## 7. Conclusion
In this paper, we have analyzed the various aspects of phishing attacks both in theory as well as in practice. We briefly presented some commonly seen attacks related to today's business world and showed some corresponding defenses to counter those attacks. We have also highlighted some recent statistical results to project the growing problem of the phishing spam. We reckon that online security threats are nothing new, and business has to cope with various types of emerging threats since the beginning of e-commerce. As business practices evolve to keep pace with the advancement of new technology, fraudulent practices also become accustomed to the new technological opportunities that present themselves. Hence, our paper proposed an anti-phishing proposal in the hope that many identity theft disasters could be avoided through creating a better understanding of the Internet frauds in general and phishing in particular.